\documentstyle[12pt]{article}
\begin{document}

\newcounter{sub}
\newcounter{subeqn}[sub]
\setcounter{sub}{\value{equation}}
\stepcounter{sub}
\renewcommand{\thesubeqn}{\alph{subeqn}}
\renewcommand{\theequation}{\thesub\thesubeqn}

\begin{titlepage}
.
\vspace{3cm}
\begin{center}
{\large\bf GAUGE INVARIANT QUANTIZATION OF DISSIPATIVE SYSTEMS OF CHARGED PARTICLES IN EXTENDED PHASE SPACE}\\ 
\vspace{3cm}
{S. KHADEMI$^{1,2}$ and S. NASIRI$^{2,3}$\\
\it$^1$ Department of Physics, Shiraz Univ., Shiraz, 71454, Iran.\\
\it$^2$ Department of Physics, Zanjan Univ., Zanjan, 313, Iran.\\ 
\it$^3$ Institute for Advanced Studies in Basic Sciences, IASBS,\\ \it P.O. Box 49195-159, Zanjan, Iran.}
\end{center}
\end{titlepage}
\newpage

{\bf Abstract-}Recently, it is shown that the extended phase space formulation of quantum
mechanics is a suitable technique for studying the quantum dissipative 
systems. Here, as a further application of this formalism, we consider a dissipative system
of charged particles interacting with an external time dependent electric 
field. Such a system has been investigated by Buch and Denman, and  two distinct
solutions with completely different structure  have been
obtained for Schr\"odinger's equation in two different gauges.
However, by generalizing the gauge transformations to the
phase space and using the extended phase space technique to study 
the same system, we demonstrate how both gauges lead to the  same conductivity,
suggesting  the recovery  of gauge invariance for this physical quantity
within the extended phase space approach.\\

{\bf Keywords-} Extended Phase Space, Electrodynamics, Gauge Transformation, Conductivity,
Quantum Dissipative System, Gauge Invariant.\\
\newpage

\begin{center}
{\bf 1. INTRODUCTION}
\end{center}
Although the gauge invariance of electrodynamics is essentially satisfied in 
theory, however, in practice, some authors have shown that the different 
gauges do not yield the same physical results.
For example, Kobe and Wen [1] have investigated an oscillating charged particle
in a time dependent electric field, in two different gauges. They have shown 
that the transition probability amplitude as an important physical quantity in
quantum mechanics, is gauge dependent. The equivalence or non equivalence of
minimally coupled and multipolar hamiltonians, which are related to each other
by a gauge transformation, is not well understood as noted by Golshan and Kobe
[2], and Cohen-Tannoudji, et. al. [3]. Noteworthy challenges about Maxwell's
equations using multipolar and minimally coupled hamiltonians are done by
Mandel [4], Healy [5], Ackerhalt [6] and others. However, their results are
not consistent with each other in all respects. Buch and Denman [7] have considered
the quantum mechanical evolution of a system of charged particles and have obteined
different structure for solutions of Schr\"odinger equation in two different
gauges.

Another point in the problem existing with the quantization of dissipative systems. Inspite of the vast efforts
done, these systems have not a well understood quantum mechanics. 
The most well-known approaches to this problem are:

a)The Kaldirolla - Kanai method [8,9], which assumes a hamiltonian for a system
of damped harmonic oscillators giving the correct classical equations of 
motions.
Although, the Kaldirolla and Kanai  hamiltonian is correct classicaly, it has
its own problem in quantum mechanics and it violates the Heisenberg uncertainty
principle.

b) The Bateman method [10], which considers a dual hamiltonian for a given 
dissipative system. The classical equations of motion of the system consider
the evolution of the system and its mirror image, simultaneously.
The energy dissipated by the original system is completely 
absorbed at the same rate by the mirror image system.  Therefore, the dual hamiltonian becomes
a constant of the motion. The difficulty existing with
this method is that the hamiltonian is not equal to the total energy of the
system and its mirror image. This presents its own hindrance to the problem.

c)The Schr\"odinger - Languvan method [11], in which a non-linear hamiltonian is 
employed in the Schr\"odinger's equation.
The outcome of this nonlinearity is, in turn, the violation of the superposition principle.

d)The Dodonov - Mankov method [12], in which the loss energy states with 
complex eigenvalues are introduced.

In a review article including considerable number of references,
Dekker [13] concludes that "...although completeness is certainly not claimed,
it is left that the present text covers a
substantial portion of the relevant work done during the last half century.
All models agree on the classical dynamics. ... closer inspection of the
models shows that none of them, ... are completely satisfactory in all
respects".

Recently, we have employed the extended phase space (EPS) formulation of quantum
mechanics [14,15,16] to investigate the evolution of dissipative systemes
[17,18,19].
We have shown that the extension of the phase space allows the presence of a 
mirror image system that absorbs the energy with the same rate
that the given system dissipates.
The whole system 
behaves as a conservative system which evolves together in the course of time 
and, therefore, the problems due to nonconservative nature of the dissipative systems
are removed.
Here, as a further application of the EPS technique, we treat the dynamics of a system of charged 
particles interacting with a time dependent electric field. To do this, 
one  needs to generalize the conventional
gauge transformations for the EPS. They are called the extended gauge transformations
and are shown to be canonical transformations in EPS.
Using the Kanai hamiltonian it is shown that, in contrast to 
the Buch and Denman treatment [7], the solutions of the evolution equation
have the same structure in two different gauges and are related to each other 
by a unitary transformation. Using these solutions the conductivity is shown 
to be a gauge invariant quantity.

The layout of this article is as follows: In section 2 a brief review of 
the EPS formulation is introduced.
In section 3 the gauge transformations are generalized for EPS and in 
section 4, by using  the proposed technique, the quantum state functions 
of the system is obteined and are used to calculate the  conductivity in two different gauges. Section 5 is
devoted to conclusions.
\begin{center}
{\bf 2. A REVIEW OF THE EPS FORMULATION}
\end{center}
A direct approach to quantum statistical mechanics is proposed by Sobouti and Nasiri [14],
by extending the conventional phase space and by applying the canonical 
quantization procedure to the extended quantities in this space.
Assuming the phase space coordinates $p$ and $q$ to be independent variables
on the virtual trajectories, allows one to define momonta $\pi_{p}$ and $\pi_{q}$,
conjugate to $p$ and $q$, respectively. This is done by introducing the
extended lagrangian

\begin{equation}
{\cal L}(p,q,{\dot p},{\dot q})=-{\dot p}q-{\dot q}p + {\cal L}^{p}(p,{\dot p})+ {\cal L}^{q}(q,{\dot q}),
\end{equation}
\stepcounter{sub}
where ${\cal L}^p $ and ${\cal L}^{q}$ are the $p$ and $q$ space lagrangians
of the given system. Using Eq. (1) one may define the momenta, conjugate to
$p$ and $q$, respectively, as follow
\setcounter{subeqn}{1}
\begin{equation}
\pi_{p} = \frac{\partial {\cal  L} }{\partial {\dot p} } = \frac{\partial {\cal L}^p}{\partial {\dot p}}-q,
\end{equation}
\stepcounter{subeqn}
\begin{equation}
\pi_{q} = \frac{\partial {\cal  L }}{\partial {\dot q} } = \frac{\partial  {\cal L} ^q}{\partial {\dot q} }-p.
\end{equation}
\stepcounter{sub}
In the EPS defined by the set of variables $\{p,q, \pi_p, \pi_q\}$, 
one may define the extended hamiltonian
\begin{eqnarray}
{\cal H}(\pi_p ,\pi_q , p,q) & = & {\dot p}\pi_p +{\dot q}\pi_q -{\cal L}= H(p+\pi_q,q)-H(p,q+\pi_p)\nonumber\\
& = & \sum\frac{1}{n!}\left\{\frac{\partial^nH}{\partial p^n}\pi_{q}^n-\frac{\partial^nH}{\partial q^n}\pi_{p}^n\right\},
\end{eqnarray}
\stepcounter{sub}
where $H(p,q)$ is the hamiltonian of the system.
Using the canonical quantization rule, the following postulates are outlined:

a) Let $p,q,\pi_p$ and $\pi_q$ be operators in a Hilbert space, $\bf X$, of
all square integrable complex functions, satisfying the following commutation
relations
\setcounter{subeqn}{1}
\begin{equation}
[\pi_{q},q]=-i\hbar,\hspace{2cm}\pi_{q}=-i\hbar\frac{\partial}{\partial q},
\end{equation}
\stepcounter{subeqn}
\begin{equation}
[\pi_{p},p]=-i\hbar,\hspace{2cm}\pi_{p}=-i\hbar\frac{\partial}{\partial p},
\end{equation}
\stepcounter{subeqn}
\begin{equation}
[p,q]=[\pi_{p},\pi_{q}]=0.
\end{equation}
\stepcounter{sub}
By virtue of Eqs. (4), the extended hamiltonian, ${\cal H}$, will also be an 
operator in ${\bf X}$.

b) A state function $\chi(p,q,t)\in{\bf X}$ is assumed to satisfy the
following dynamical equation 
\begin{eqnarray}
i\hbar\frac{\partial \chi}{\partial t}&=&{\cal H}\chi=[H(p-i\hbar\frac{\partial}{\partial q},q)
-H(p,q-i\hbar\frac{\partial}{\partial p})]\chi\nonumber\\
&=&\sum\frac{(-i\hbar)^n}{n!}\left\{\frac{\partial^nH}{\partial p^n}\frac{\partial^n}{\partial q^n}-\frac{\partial^nH}{\partial q^n}
\frac{\partial^n}{\partial p^n}\right\}\chi.
\end{eqnarray}
\stepcounter{sub}
c) The averaging rule for an observable $O(p,q)$, a c-number operator in this 
formalism, is given as
\begin{equation}
<O(p,q)>=\int O(p,q)\chi^{*}(p,q,t)dpdq.
\end{equation}
\stepcounter{sub}
For details of selection procedure of the admissible state functions, see 
Sobouti and Nasiri [14].

\begin{center}
{\bf 3. GENERALIZATION OF  GAUGE TRANSFORMATIONS FOR EPS}
\end{center}
Interaction of a charged particle with an external electromagnetic field is
described by the following hamiltonian:
\begin{equation}
H=\frac{1}{2m}[p-\frac{e}{c}A]^2+e\phi,
\end{equation}
\stepcounter{sub}
where $A(q,t)$ and $\phi(q,t)$ are electromagnetic vector and scalar potentials
and $e$ is the electric charge of the particle.
The conventional gauge transformations are unitary transformations as follows
\begin{equation}
F=\exp\left(\frac{-ie}{\hbar c}f(q,t)\right),
\end{equation}
\stepcounter{sub}
where $f(q,t)$ is an arbitrary gauge function in $q$-space [20]. 
Shr\"odinger's equation in q-representation would be form invariant under
the above transformation. This requires that $A$ and $\phi$
transform as follows [20]
\setcounter{subeqn}{1}
\begin{equation}
A^{\prime}=A+\nabla_{q} f(q,t),
\end{equation}
\stepcounter{subeqn}
and
\begin{equation}
\phi^{\prime}=\phi -\frac{1}{c}\frac{\partial f(q,t)}{\partial t}.
\end{equation}
\stepcounter{sub}
Alternatively, this gauge invarianc may be obtained by
assuming $H\rightarrow H-\frac{e}{c}\frac{\partial f(q,t)}{\partial t}$,
and $ p\rightarrow p-\frac{e}{c}\nabla f(q,t)$.
This  gives, $p-\frac{e}{c}A\rightarrow (p-\frac{e}{c}\nabla f)-\frac{e}{c}A=p-\frac{e}{c}(A+\nabla f)$.
Thus, one may consider the term $\frac{e}{c}\nabla f $ either with
$p$ or  with $A$.
So  the  ordinary gauge trasnsformations may be looked either as a gauged
transformation on potentials or a canonical coordinate transformation on
the phase space coordinates $p$ and $q$ with  fixed  $q$. The later
interpretation  will be important in the  definition of gauge transformations
in the extended phase space.

Using Eq. (3), one may obtain the extended hamiltonian
for a charged particle interacting with an extertnal electromagnetic field,
as follows
\begin{eqnarray}
{\cal H}&=&\frac{1}{2m}[\pi_{q}-\frac{e}{c}A(q,t)]^2 +\frac{1}{m}p.[\pi_{q}-\frac{e}{c}A(q,t)]+e\phi(q,t)\nonumber\\
& &-\frac{1}{2m}[\frac{e}{c}A(q+\pi_{p},t)]^2 -\frac{1}{m}[\frac{e}{c}p.A(q+\pi_{p},t)]
-e\phi(q+\pi_{p},t).
\end{eqnarray}
\stepcounter{sub}
We assume that the gauge transformation in EPS may be described by a unitary operator 
as follows
\begin{equation}
\Gamma=\exp\left(\frac{-ie}{\hbar c}\gamma(q,p,t)\right),
\end{equation}
\stepcounter{sub}
where $\gamma(q,p,t)$ is now an arbitray gauge function of the phase space 
coordinates. In order  that Eq. (5) to be form invariat under 
transformation Eq. (11),
one requires

\setcounter{subeqn}{1}

\begin{equation}
\chi^{\prime}=\Gamma\chi,
\end{equation}

\stepcounter{subeqn}

\begin{equation}
{\cal H}^{\prime}=\Gamma{\cal H}\Gamma^{\dagger}-i\hbar\Gamma\left(\frac{\partial \Gamma^{\dagger}}{\partial t}\right),
\end{equation}

\stepcounter{sub}
where $\chi^{\prime}$ and ${\cal H}^{\prime}$ are gauge transformed state 
function and extended hamiltonian, repectively.
To be consistent with conventional gauge transformations in 
$q$-representation, one requires the following form for $\Gamma$ 
\setcounter{subeqn}{1}
\begin{equation}
\Gamma=FG^{\dagger},
\end{equation}
\stepcounter{subeqn}
where 
\begin{equation}
G=\exp\left(\frac{-ie}{\hbar c}g(p,t)\right).
\end{equation}
\stepcounter{sub}
In Eq. (13b), $g(p,t)$ is an arbitrary gauge function in  p-space.
In fact, the unitary transformation $G$ leaves Schr\"odinger's equation form invariant in p-representation 
in parallel with the $F$ which does the same job for q-representation.
Eventually, using Eq. (12b), the gauge transformed extended hamiltonian will
 have the following form 
\begin{eqnarray}
{\cal H}^{\prime}& = &(FH(p+\pi_{q},q)F^{\dagger}-i\hbar F\frac{\partial F^{\dagger}}{\partial t})-(G^{\dagger}H(p,q+\pi_{p})G-i\hbar G^{\dagger}\frac{\partial G}{\partial t})\nonumber\\
& = &\frac{1}{2m}[\pi_{q}-\frac{e}{c}\nabla_{q}f(q,t)-\frac{e}{c}A(q,t)]^2+\frac{1}{m}p.[\pi_{q}-\frac{e}{c}\nabla_{q}f(q,t)-\frac{e}{c}A(q,t)]+e\phi(q,t)\nonumber\\
& &-\frac{1}{2m}[\frac{e}{c}A(q+\pi_{p}+\frac{e}{c}\nabla_{p}g(p,t))]^2
-\frac{e}{mc}p.A(q+\pi_{p}+\frac{e}{c}\nabla_{p}g(p,t))\nonumber\\
& &-e\phi(q+\pi_{p}+\frac{e}{c}\nabla_{p}g(p,t),t).
\end{eqnarray}
\stepcounter{sub}
Note that, the arguments of $A$ and $\phi$ in the last three terms of ${\cal H}^{\prime}$ 
in Eq. (14), does not allow to express the extended gauge transformation 
in their standard form, that is, in terms of vector and scalar potentials
as in Eq.(9). In fact  the additive property of $A$ and $\phi$ with derivative
of gauge functions is lost.
However, another possibility emerges. One may consider the  extended gauge
transformations as the following coordinate transformations
\setcounter{subeqn}{1}
\begin{equation}
\pi_{p}^{\prime}=\pi_{p}+\frac{e}{c}\nabla_{p}g(p,t),\hspace{2cm}p^{\prime}=p,
\end{equation}
\stepcounter{subeqn}
\begin{equation}
\pi_{q}^{\prime}=\pi_{q}-\frac{e}{c}\nabla_{q}f(q,t),\hspace{2cm}q^{\prime}=q,
\end{equation}
\stepcounter{sub}
where $\nabla_{p}$ denotes the derivative with respect to $p$.
It is simply verified that the Eqs. (15) in classical level, are canonical coordinate 
transformations which in quantum level correspondes to unitary transformations.
Note that the interpretation of gauge transformations as canonical coordinate
transformations is valid in both of the ordinary phase space
as well as extended phase space. While the usual form of the gauge transformations
on the electromagnetic potentials are valid only in  ordinary phase space
as noted earlier.

\begin{center}
{\bf 4. APPLICATION TO A SYSTEM OF DISSIPATIVE CHARGED PARTICLRES}
\end{center}
Let us now apply this method, to study a dissipative system of non interacting charged
particles in a time dependent uniform external electric field.
The single particle hamiltonian for this medium with damping constant
$\alpha$ is as follows [9]
\begin{equation}
H=\frac{1}{2m}e^{-\alpha t}[p-\frac{e}{c}A(q,t)]^2 +e^{\alpha t}e\phi(q,t).
\end{equation}
\stepcounter{sub}
Extending of Eq. (16), one gets 
\begin{eqnarray}
{\cal H}(p,q,\pi_{p},\pi_{q})&=&\frac{1}{2m}e^{-\alpha t}[p+\pi_{q}-\frac{e}{c}A(p,q)]^2+e^{\alpha t}e\phi(q,t)\nonumber\\
& &-\frac{1}{2}e^{-\alpha t}[p-\frac{e}{c}A(q+\pi_{p},t)]^2 -e^{\alpha t}e\phi(q+\pi_{p},t).
\end{eqnarray}
\stepcounter{sub}
Using Eq. (17), we investigate the behavior of the system in two different gauges; 
one with zero scalar potential and the other with zero vector potential
(hereafter called $A$-gauge and $\phi$-gauge, respectively).
This is done in the following subsections.

\begin{center}
{\bf 4.1. ${\bf A}$-GAUGE}
\end{center}
A time dependent uniform electric field may be generated by setting
\begin{equation}
A(t)=-c\int^t e^{\alpha \lambda}E(\lambda)d\lambda ,\hspace{2cm}\phi(q,t)=0,
\end{equation}
\stepcounter{sub}
which is called A-gauge. Note that $A(t)$ depends only on time. In this gauge the
extended hamiltonian (17), assumes the following form
\begin{equation}
{\cal H}_{A}=\frac{1}{2m}e^{-\alpha t}[\pi^2_q +2\pi_q(p+e\int^t e^{\alpha \lambda}
E(\lambda)d\lambda)].
\end{equation}
\stepcounter{sub}
In obtaining Eq. (19), the gauge transformation was followed by the extension.
A word of caution is in order. These operations do not, in general, commute
with each other and have its own interesting consequences that will be presented
elsewhere. Of course the electromagnetic field assumed here is a special case, hence, the ordering
of operations concerning with extension and gauge transformation is not important.
Using the above hamiltonian, the evolution equation [15], becomes
\begin{equation}
i\hbar\frac{\partial \chi_{_A}}{\partial t}={\cal H}_{A} \chi_{_A}=
e^{-\alpha t}\left[-\frac{\hbar^2}{2m}\frac{\partial^2 \chi_{_A}}{\partial q^2}-
\frac{i\hbar}{m}\left(p+e\int^t e^{\alpha \lambda}E(\lambda)d\lambda\right)\frac{\partial \chi_{_A}}{\partial q} \right].
\end{equation}
\stepcounter{sub}
Making the transformation 
\setcounter{subeqn}{1}
\begin{equation}
\xi=q-\frac{p}{m}\int^t e^{-\alpha \lambda}d\lambda-\frac{e}{m}\int^t e^{-\alpha \lambda}
\int^{\lambda} e^{\alpha \nu}E(\nu)d\nu d\lambda,
\end{equation}
\stepcounter{subeqn}
\begin{equation}
\eta=p,
\end{equation}
\stepcounter{subeqn}
\begin{equation}
\tau=t,
\end{equation}
\stepcounter{sub}
and using them in Eq. (20), one gets
\begin{equation}
i\hbar\frac{\partial \chi_{_A}}{\partial \tau}=-\frac{\hbar^2}{2m}e^{-\alpha \tau}\frac{\partial^2 \chi_{_A}}{\partial \xi^2}.
\end{equation}
\stepcounter{sub}
Equation (22) can be solved for $\chi_{_A}$ by separation of variables and 
the result is 
\begin{equation}
\chi_{A}=C_{+}\exp\left[+ik\xi +\frac{i\hbar k^2}{2m\alpha}e^{-\alpha\tau}\right]
+C_{-}\exp\left[-ik\xi+\frac{i\hbar k^2}{2m\alpha}e^{-\alpha \tau}\right],
\end{equation}
\stepcounter{sub}
where $C_{\pm}$ and $k$ are normalization and separation constants, respectively. 
One can use the state function $\chi_{_A}$ of Eq. (23) to calculate the conductivity
for the system of $N$ non interacting charged particles given by 
\begin{equation}
\sigma=\frac{Ne<{\dot q}>}{E(t)},
\end{equation}
\stepcounter{sub}
where 
\begin{equation}
{\dot q}=\frac{1}{m}pe^{-\alpha t}+\frac{e}{m}e^{-\alpha t}\int^t E(\lambda)e^{\alpha \lambda}d\lambda.
\end{equation}
\stepcounter{sub}
The expression for ${\dot q}$ in Eq. (25) can be obtained using extended
Hamilton's equatios as $\pi_{q}\rightarrow 0$ [14].
If one sets $C_{+}=C_{-}$ and calculates the expectation value of $<{\dot q}>$
using the averaging rule of Eq. (6), one gets
\begin{equation}
\sigma=\frac{Ne^2}{m(\alpha +i\omega)},
\end{equation}
\stepcounter{sub}
as $<p>$ vanishes. A time dependency of the form $E(t)=E_{0}e^{i\omega t}$
is assumed for the external electric field. Equation (26) is consistent with 
the result obtained by  Buch and Denman [7] in the same gauge.

\begin{center}
{\bf 4.2. $\bf\phi$-GAUGE} 
\end{center}
The aformentioned electric field may be obtiened by assuming 
\begin{equation}
A(q,t)=0,\hspace{2cm}\phi(q,t)=-qE(t).
\end{equation}
\stepcounter{sub}
The corresponding extended hamiltonian will be 
\begin{equation}
{\cal H}_{\phi}=e^{-\alpha t}\frac{1}{2m}\left(\pi_{q}^2+2p\pi_{q}\right)
+e^{\alpha t}eE(t)\pi_{p}.
\end{equation}
\stepcounter{sub}
The evolution equation now becomes 
\begin{equation}
i\hbar\frac{\partial \chi_{\phi}}{\partial t}={\cal H}_{\phi}\chi_{\phi}=
e^{-\alpha t}\left(-\frac{\hbar^2}{2m}\frac{\partial^2 \chi_{\phi}}{\partial q^2}-
\frac{i\hbar p}{m}\frac{\partial \chi_{\phi}}{\partial q}\right)-i\hbar e^{\alpha t}e
E(t)\frac{\partial \chi_{\phi}}{\partial p}.
\end{equation}
\stepcounter{sub}
In parallel with Eqs. (21), one may consider the following transformation 
\setcounter{subeqn}{1}
\begin{equation}
\xi^{\prime}=q-\frac{p}{m}\int^t e^{-\alpha \lambda}d\lambda+\frac{e}{m}\int^t e^{\alpha \lambda}
E(\lambda)\int^{\lambda} e^{-\alpha \nu}d\nu d\lambda,
\end{equation}
\stepcounter{subeqn}
\begin{equation}
\eta^{\prime}=p-e\int^te^{\alpha \lambda}E(\lambda)d\lambda.
\end{equation}
\stepcounter{subeqn}
\begin{equation}
\tau^{\prime}=t.
\end{equation}
\stepcounter{sub}
Equation (29) under the above transformation becomes
\begin{equation}
i\hbar\frac{\partial \chi_{\phi}}{\partial \tau^{\prime}}=-\frac{\hbar^2}{2m}e^{-\alpha \tau^{\prime}}\frac{\partial^2 \chi_{\phi}}{\partial \xi^{\prime 2}}.
\end{equation}
\stepcounter{sub}
The remarkable point is that, except for different functional forms of $\xi$
and $\xi^{\prime}$, Eq. (31) has exactly the same form as Eq. (22). 
Therefore, we conclude that, in contrast to the result obtained by Buch and 
Denman using the conventional Schr\"odinger quantum mechanics, the solution
$\chi_{\phi}$ of Eq. (31) has the same structure as $\chi_{_A}$ of Eq. (22)
and differ only by a phase factor.
This result may be considered as an advantage of the EPS method.
In fact, $\chi_{\phi}$ and $\chi_{_A}$ are 
related to each other by a unitary transformation, and therefore, one can easily
 verify that the value obtained for electrical 
conductivity in $A$-gauge will be valid for $\phi$-gauge, too.

\begin{center}
{\bf 5. CONCLUSIONS}
\end{center}
The gauge transformations in electrodynamics are ganeralized to the extended 
phase space and is shown to have their own advantages in studying the quantum
mechanical evolution of charged particles. They turned out to be canonical 
dcoordinate transformations in this space.
The interpretation of the gauge transformations as change of electromagnetic
poientials is no longer valid in EPS approach. However, treating the gauge transformations
as canonical
coordinate transformations is valid for EPS method as well as for
conventional approachs. In other words, it seems more convenient and 
resonable to treat the gauge transformations as canonical coordinate transformations.
The formalism is applied 
to study the evolution of a system of charged particles in
the presence of a uniform and time dependent external electric field. 
The problem is handled in two different gauges,
first employed by Buch and Denman to study the same problem by using the 
Schr\"odinger quantum mechanics. They obtained two physically different solutions
for Schr\"odinger's equation using above gauges.
However, our solutions, in contrast to those of Buch and Denman,
have a unique structure in both gauges for any arbitrary boundary conditions.
They are related by a unitary transformation which garrantee the gauge equivalence
of the physical quantities.
As an example, the conductivity is calculated for the system and is shown that it
is a gauge independent quantity.

There is also a noteworthy point in obtaining Eqs. (19) and (28). In general, the 
operations of extension and gauge transformations do not commute with each other.
It depends on the form of hamiltonian and the 
interacting electromagnetic fields. However, for the special case considered here, no 
mattor how these operations might be ordered. The more general case is investigated
and the results will be appeared elsewhere.

\begin{center}
{\bf ACKNOWLEDGMENT}\\
\end{center}
We would like  to thank  Prof. Y. Sobuti for his helpful comments. This work is 
supported by Zanjan University. 

\begin{center}
{\bf REFRENCES}
\end{center}
1. Kobe, D., H., and Wen, E., C., T., Phys. Letter, 80 A, 121 (1980).\\
2. Golshan, S., M., and Kobe, D., H., Phys. Rev. A, 34, 4449 (1986).\\
3. Cohen-Tannoudji, C., Dupont-Roc, J., and Grynberg,G., {\it Photons and Atoms},
   John Wiley \& Sons. Inc., (1989).\\
4. Mandel, L., Phys. Rev. A, 20, 1590 (1979).\\
5. Healy, W., P., Phys. Rev. A, 26, 1798 (1982).\\
6. Ackerhalt, J., R., and Milloni, P., W., J. Opt. Soc. Am. B, 1, 116 (1984).\\
7. Buch, L. H., and Denman, H. H., Am. J. Phys., 42, 304 (1974).\\
8. Kaldirolla, P., Nuovo Cimento, 18, 393 (1941).\\
9. Kanai, E., Prog. Theo. Phys., 3, 440 (1948).\\
10. Bateman, H., Phys. Rev., 38, 393 (1931).\\
11. Razavy, M., Can. J. Phys., 56, 311 (1978).\\
12. Dodanov, V., and Mankov, V., Nuovo Cimento, 403 (1978).\\
13. Dekker, H., Phys. Rep., 80, 1 (1981).\\
14. Sobouti, Y., and Nasiri, S., Int. J. Mod. Phys., B, 18, 7 (1993).\\
15. Sobouti, Y., and Nasiri, S., Tr. J. Phys., 19,2 (1995).\\
16. Nasiri, S., Iranian J. Sci. Tech., 19,2 (1995).\\
17. Nasiri, S., (Submitted to Iranian J. Phys.) (2000).\\
18. Nasiri, S., Pahlavani, H., and Fallahi, M. T.,(Submitted to Iranian J. Phys.) (2000).\\
19. Nasiri, S., and Nafari, N., (Submitted to Journal of Sciences, JSIRI) (2000).\\ 
20. Kobe, D., H., and Smirl, A., L., Am. J. Phys. 46, 624 (1978)
21. Kobe, D., H., Am. J. Phys., 56, 252 (1983).\\
22. Power, E., A., and Thirunamachandran, T., Phys. Rev. A, 28, 2649 (1983).\\
23. Sipe, J., E., Phys. Rev. A, 27, 615 (1983).
\end{document}